\documentstyle[aps,prl,epsfig,twocolumn]{revtex}
\newcommand{\beeq}{\begin{equation}}
\newcommand{\eneq}{\end{equation}}
\newcommand{\beeqar}{\begin{eqnarray}}
\newcommand{\eneqar}{\end{eqnarray}}
\begin{document}
\twocolumn[\hsize\textwidth\columnwidth\hsize\csname
  @twocolumnfalse\endcsname
\preprint{IMSc 97/07/31}
\title{An Asymptotically Free Extension of QCD and ALEPH Four Jet Events}
\author{N.D. Hari Dass \dag} 
\address{The Institute of Mathematical Sciences, Chennai - 600 113, INDIA}
\author{V. Soni\ddag} 
\address{National Physical Laboratory, K.S. Krishnan Road, New
Delhi, INDIA}
\maketitle
\begin{abstract}
We show that a recently proposed extension of QCD by the addition of 
a multiplet of scalars transforming as (2,2,1) under $SU(2)_L\times 
SU(2)_R\times SU(3)_{colour}$
provides a natural and parameter-free explanation for many observed 
features of the excess 4 jet events reported by ALEPH.
\end{abstract}
\vskip 2pc] 
We have recently proposed a theory that is an extension of QCD 
by the addition of
a multiplet of elementary scalars $(\tilde\sigma,\vec{\tilde\pi}$) transforming as (2,2) 
under an $SU(2)_L\times
SU(2)_R$ and as a  singlet under the colour group $SU(3)$ and interacting only 
with quarks.
We showed that such a theory  could have an 
asymptotically free(AF) phase , where all couplings including the yukawa 
and scalar
self couplings are AF, if the initial value of the the ratio 
of the yukawa
coupling,  $g_y$ , and the QCD coupling , $g_3$ , was smaller than 
a critical value \cite{soni}.

A version of  this theory with the chiral scalars having light masses 
had all the properties of conventional
QCD and in fact only the high-precision Z- width($\Gamma_Z$) data could favour QCD
over this theory\cite{ztest} .

The other possible version of this theory where chiral
symmetry is manifest in the scalar sector, and the scalar mass is 
$\ge $ 45 Gev
is without any conflict  with 
known data.

In such a theory, 
a distinctive signal will be the appearance of an excess of
four jet events in $e^+ e^-$ - collisions over what is to be expected 
from the standard model.
Such four jet signals arise as the massive scalars eventually decay into
$\bar q q$ pairs.

Recently, the ALEPH collaboration \cite{aleph1} has reported seeing such an excess of
four jet events in $e^+ e^-$ collisions at $\sqrt s = $ 130,136,161 and
172 Gev respectively.
It was reported that there was a preponderance
of light flavours in the jets and also no associated leptons were
reported.
Both these features are significant because extensions of the standard
model available in the literature like `two Higgs Doublet models'
(THDM), and supersymmetric extensions like SSM, MSSM etc
generically predict dominant branching ratios for heavy flavours as well
as associated leptons.In contrast, our model requires more or less
uniform coupling to all flavours if significant flavour changing 
neutral currents are to be avoided, and naturally avoids events with
associated leptons.

At the same time none of the precision tests of the standard model
including the so called oblique parameters S,T\&U  or the branching ratio of Z
into
$b\bar b$ are in conflict with our model.In this letter we present the salient 
features of our model and discuss the ALEPH four jet events in its light.

{\bf The Model}:
our model is described by
the lagrangean 
\begin{eqnarray}
{\cal L} ={\cal L}_{QCD} & - & \frac{1}{2} [(\partial_\mu \tilde\sigma)^2 
+(\partial_\mu \vec{\tilde\pi})^2]
-\lambda {(\tilde\sigma^2+\vec{\tilde\pi}^2)}^2\nonumber \\
& - & \overline\Psi_q\left[
g_y(\tilde\sigma+i\gamma_5 \vec\tau\cdot\vec{\tilde\pi})\right ]\Psi_q
+{M_{\pi}^2\over 2}(\tilde\sigma^2+\vec{\tilde\pi}^2)
\label{lagan}
\end{eqnarray}
$g_y$, and  $\lambda$ are the Yukawa,and scalar self- couplings respectively.
 $\overline\Psi_q$ is the quark field. 
 
 A model described by
 eqn(1) 
 automatically implies that the chiral multiplet couples to the
 electroweak gauge bosons. 
Representing the chiral multiplet as a complex doublet
$\Phi_{ch}^{T} = (\tilde\sigma+i\tilde\pi,i\sqrt{2}\tilde\pi{-})$,
the minimal coupling to the electroweak gauge bosons is
given by 
\beeq
{\cal L}_{gauge,chiral} = {1\over 2}|(\partial_\mu-i{g\over 2}
\vec\tau\cdot \vec A_{\mu}+i{g^{\prime}\over 2}B_{\mu})\Phi^{ch}|^2
\eneq
For example, the linear couplings that result are
\beeqar
{\cal L}_{lin}^{neut}&=& e(A_{\mu}-{\gamma\over 2} Z_{\mu})
(\vec{\tilde\pi}\times
\partial_\mu\vec{\tilde\pi})_3
-{e\over 2cs}Z_\mu(\tilde\pi_0\partial_\mu\tilde\sigma-\tilde\sigma
\partial_\mu\tilde\pi_0)
\nonumber\\
{\cal L}_{lin}^{char}&=& {g\over 2}W_-^{\mu}[(\tilde\pi_+\partial_\mu
\tilde\sigma-\tilde\sigma
\partial_\mu\tilde\pi_+)+i(\tilde\pi_+\partial_\mu\tilde\pi_0
-\tilde\pi_0\partial_\mu\tilde\pi_+)]\nonumber\\
& &+c.c
\eneqar
where $\gamma = (1-2s^2)/cs)$ with $s= sin \theta_W$ and $c = cos
\theta_W$. 

As shown in \cite {ztest}, this model with
zero or very light mass scalars is incompatible with precision data on 
$\Gamma_Z$. However,
when chiral symmetry is manifest with the degenerate
scalar mass
 $M_{\tilde\pi} \ge
 M_Z/2$, the model is not ruled out by $\Gamma_Z$ data.
The mass term in eqn(1) can arise from an
underlying Higgs effect from
$ {\cal L}_{Higgs}^{chiral} = \eta |\Phi^{ch}|^2|H|^2$
where H is the Higgs-doublet of the standard model.
Renormalisability of the 
theory would also require this.

{\bf The RNG Flows}:
the $\beta$ function for the QCD coupling, $\alpha_s=g_3^2$, is
unchanged at one loop level.
The yukawa coupling $g_y$ has the following one-loop $\beta$
function (assuming 3 colours, 6 flavours and ignoring the contributions from the
standard model gauge couplings) 
\begin{equation}
\frac{\partial {g_y^2}}{\partial t} = \frac{g_y^2}{8\tilde\pi^2}
\left [ 36 g_y^2 - 8 \alpha_s \right ]
\label{dgy}
\end{equation}
The chiral multiplet 
couples to all
generations of quarks.  

The detailed analysis of these flow equations will be presented
elsewhere \cite{rng}. The result relevant to us here is that there is a regime
 $0< \rho<1/36$(here  
$\rho=g_y^2/\alpha_s$) where $g_y$ is asymptotically free with the deep
asymptotic behaviour 
 $g_y^2 \sim K\alpha^{8/7}$.
This means that $g_y^2$  vanishes faster than 
$\alpha_s$. Therefore, the leading behaviour of this theory in the 
ultraviolet is
given by the QCD coupling with the yukawa coupling contributing only in 
sub- leading order. 
There is a whole family of
solutions corresponding to different K's . \par

The above analysis is valid for $q^2 \ge m_t^2$. For the region
$m_b \le q \le m_t$, one has: $\rho_c = 1/108$,
$\rho \simeq K\alpha ^{1/23}$. 
For $N_F \le
4$,eqn(4) has no 
AF solution. Inclusion of contributions due to the
standard model gauge couplings,however, improves the situation in that
$\rho_c$ is increased and AF persists
for all values of $N_F$\cite{rng}.The self-coupling $\lambda$ also admits 
AF flows.

Thus we have classes of theories that are not only asymptotically free
in all their couplings,
 but become increasingly indistinguishable from QCD at high energies. 
As far as AF is concerned 
one loop analysis is stable against higher loop corrections 
\cite{gross}. 
Since these classes of theories are AF with chiral symmetry, they are all candidates for a 
consistent theory of strong interactions.

In order to apply these results to energy scales of practical interest,
threshold effects associated with masses of various particles
have to be taken into account.This will be done elsewhere\cite{rng}.
Fortunately, for the analysis presented in this paper, these details
are not needed.

By construction, there is no coupling between the
leptons and the chiral scalars. The reasons are two fold: i) the lepton
sector does not have any non-trivial $SU(2)_L\times SU(2)_R$ structure
in any limit,ii) unlike the
coupling of the chiral scalars to the electroweak bosons, nothing
demands that the scalars couple to the leptons.
This is a very important distinction between our
model and extensions of the standard
model like THDM,SSM,MSSM
etc.

{\bf Flavour Changing Neutral Currents}:
In the standard model, $\Delta F = 2$ processes mediated
at one loop level are severely suppressed.
It is important that significant FCNC are not introduced in our model.
In order that interactions preserve $SU(2)_L$- invariance, one must have
\beeq
{\cal L}_{yukawa} =
 F_{AB}\bar \Psi_A^{\prime} (\tilde\sigma + i\gamma_5 \vec
\tau\cdot \vec{\tilde\pi})\Psi_B^{\prime}
\eneq
Here
$\Psi^{\prime}_A$
is the Dirac field in the gauge - eigenstate basis. 
There are potential FCNC terms even at tree level in eqn(5)
but the mass degeneracy of 
$\tilde\sigma$ and $\tilde\pi^0$ leads to a complete cancellation.
This is true both for 
$\Delta F = 2 $ and $ \Delta F = 1 $ processes.
This cancellation 
persists even for the one-loop contributions.
This is unlike the
situation
in generic two Higgs doublet models where the couplings have to
be fine-tuned to keep tree-level FCNC under control.
In such models,the
analogs of $\tilde\sigma$ and $\tilde\pi^0$ are not mass-degenerate.
The detailed analysis of the FCNC problem will be
presented elsewhere \cite{fcnc}. Here we shall only quote the main result:
{\bf F must be a multiple of unit matrix} and the interaction in
the mass-eigenstate basis is given by 
\beeqar
{\cal L}_{yukawa} 
& =&g_y[ \bar p_LW_d^1 (\tilde\sigma + i\tilde\pi^0)p_R+
 \bar n_LW_d^2 (\tilde\sigma - i\tilde\pi^0)n_R\nonumber\\
&+ &i\tilde\pi^- \bar n_LV_{CKM}W_d^1  p_R+
i\tilde\pi^+ \bar p_LV_{CKM}W_d^2  n_R]
\eneqar
where $W_d^{1,2}$ are diagonal unitary matrices and $V_{CKM}$ the CKM-matrix.

In the absence of spontaneous symmetry breaking of $SU(2)_L\times
U(1)$ and the limit of no $U(1)$-coupling, our
solution enhances the symmetry in the quark sector from $SU(2)_L\times SU(2)_R$
to 
$SU(3)_{hor}\times SU(2)_L\times SU(2)_R$ and hence does not amount to 
fine-tuning. 
Minimal CP-nonconservation can be
achieved by taking $W_d^{(1,2)}$ to be unit matrices, in which case the CKM matrix again 
governs the entire CP-violation in this model too.

{\bf Unitarity Constraints}:
one of the key properties of the standard model is the unitarity of all
scattering amplitudes. The standard model Higgs 
kills the bad high-energy
behaviour of processes like $WW \rightarrow WW$ and $WW \rightarrow
\bar f f$ where $f$ is any fermion. Introduction of new bosons
into
the theory that couple to fermions and gauge bosons should 
not spoil this . The necessary and
sufficient conditions for this are \cite {haber}
\beeqar
\Sigma_i g_{h^0_i V V}^2 & = & g_{hVV}^2\nonumber\\
\Sigma_i g_{h^0_i V V} \cdot g_{h^0_i f \bar f}& = & g_{hVV}\cdot
g_{hf\bar f}
\eneqar
where h is the standard model Higgs and $h_i^0$ are all the neutral
scalar fields of the theory including the analog of h. In our theory,
since there is no spontaneous symmetry breaking in the chiral - multiplet sector, no $\chi VV $
coupling is introduced and the above conditions are trivially
satisfied.

{\bf Other Precision Tests for the Model}:
the coupling of the chiral multiplet to the electroweak bosons
leads to additional contributions to various 
processes of the electroweak theory.One of the sensitive tests for QCD
is the value of the R-parameter.As the cross-section for the pair
production of the scalars at LEP energies is of the order of a 1pb, the
R-parameter is not very sensitive to the presence of the scalars. The
other important precision test is the g-2 for muons.
There are two types of additional contributions
to g-2 that arise. One is due to the enhanced ultraviolet degrees of
freedom
and the other due to  additional hadronic
interactions . The contributions due to the former 
arise out of the
modification of the photon propagation function 
The analytic result is 
$\Delta g = {\alpha\over 2\pi}{\alpha\over 180\pi}{m_{\mu}^2\over m_{\tilde\pi}^2}$
For $m_{\tilde\pi} = 45 GeV$ this amounts to $\Delta g = 4\cdot 10^{-14}$ and hence
insignificant. The shift due to the modification of hadronic interactions is
much harder to estimate precisely.  
One should expect very little difference between QCD and the 
extended theories here because of the expected decoupling of massive
particles. 
One expects the additional interactions
to produce changes in g-2 at the level of not more than  ${g_y^2 \over 4\pi} {1\over 2\pi}$
times the dominant hadronic contributions. This amounts to less than 2 parts in
 1000 of the dominant hadronic contributions and is hence much less than the
known theoretical uncertainties in g-2. 

The so called oblique parameters $\tilde S,\tilde T,\tilde U$ 
(see for example \cite {hagi})
can also
be used as 
precision tests for our model.
We skip the details of their calculations which are presented 
elsewhere \cite{stu} and give here the corrections to
these parameters:
\vspace{0.1in}
\begin{center}
\begin{tabular}{llll}
\hline
\hline\\
Scalar Mass(GeV) & $~~~  \delta\tilde T$& $~~~    \delta\tilde S$ & 
$~~~    \delta\tilde U$ \\ 
\hline
\hline\\
 ~~~~~~~~     50 &-.010  & -.027  & .005   \\ 
 ~~~~~~~~     55 &-.005  & -.018  & .002   \\ 
 ~~~~~~~~     60 &-.003  & -.013  & .001   \\ 
\hline
\hline\\
\end{tabular}
\end{center}
these corrections are much smaller than the uncertainties in even the
most precise LEP measurements \cite{hagi}

{\bf The Aleph Experiment}:
the ALEPH collaboration has recently reported seeing an excess of four
jet events in $e^+e^-$-collisions at LEP.
At $\sqrt s= $ 130 and 136 Gev they had reported seeing 16 four jet events (after
imposing various cuts; for details see \cite{aleph1})while only 8.6
events were expected from the standard model.The sum of the di-jet masses
was observed to peak  around 105 Gev, while the difference in
dijet masses peaked at low values but not exactly at zero.Subsequently,
the initial runs carried out at $\sqrt s = $
161 Gev with an integrated luminosity of about 2.5$pb^{-1}$ failed to
confirm the excess seen earlier.The status of the runs at 161 Gev with
increased data collection (10 $pb^{-1}$) was that while the excess 
could not be
confirmed., it could not be ruled out either \cite {rag}.

After completing some preliminary runs at $\sqrt
s = $ 172 Gev with roughly 7 $pb^{-1}$ of data, the ALEPH collaboration
has confirmed seeing the excess of four jet events again \cite{rag}. They report
18 events as against 3 expected, with the di-jet 
mass sum peaking at 106 Gev, not very different from what was seen at
130-136 Gev.

The initial runs with $\sqrt s= $130 and 136 Gev with an integrated
luminosity of $\sim $5.7 $pb^{-1}$ corresponded to a cross-section of
3.1$\pm$1.7 pb \cite{aleph1} with certain assumptions regarding efficiencies.The
integrated data 130/172 Gev yields a somewhat smaller cross-section
of 2.5$\pm$0.7pb under the same assumptions about efficiencies,while
the higher energy runs 161/172 Gev yield 1.5$\pm$0.8 pb \cite{rag}.

The dijet mass sum peaked at 103-109 GeV with a
resolution 1.6-9 GeV \cite{aleph1}. The global fit including the higher 
energy runs
yields the peak at 106.1$\pm$0.8 GeV with a resolution of 2.1$\pm$0.4
Gev \cite{rag}. 
For the initial runs at 130,136 GeV the following additional features
were reported \cite{aleph1}: 
a) no lepton with high transverse momentum wrt its
jet was found.
b) only 1 out of the 12 peak events could be identified
as having at least 2 b($\bar b$) jets
c) no event can be identified as having all four b($\bar b$) jets
d) the number of $K_s^0$ found was compatible with a normal flavour mixture 
of four-jet events
and e) no events of the type $\tau^+\nu_{\tau}\bar cs$ and 
$\tau^+\nu_{\tau}\tau^-\bar\nu_{\tau}$ were found.

In our model the excess events are identified with the decay products of the
scalars which have no couplings to the leptons naturally explaining
the points a) and e). The following table gives in pb the cross-sections
for $e^+e^-\rightarrow \tilde\sigma\tilde\pi^0 (\sigma_{neut})$ and to 
$\tilde\pi^+\tilde\pi^-(\sigma_{char})$ as a function of $\sqrt s$ and scalar mass.Taking 
into account the FCNC constraint that the scalars couple nearly equally
to all flavours, the branching ratios for having at least 2 b($\bar b$)
jets ($R_{2b}$) and four b($\bar b$) jets ($R_{4b}$) are also given. 

\begin{center}
\begin{tabular}{llllll}
\hline
\hline\\
$\sqrt s$(GeV) & $~~~  M_{\tilde\pi}$& $~~~ \sigma_{neut}$ & $~~~   \sigma_{char}$
&~~~~$R_{2b}$&$~~~~R_{4b}$ \\ 
\hline
\hline\\
 ~~~~~130 &~~~50  &~~~ .63  &~~~ .54&~~~1/5&~~~1/46   \\ 
 ~~~~~130 &~~~55  &~~~ .37  &~~~ .31 &~~~1/5&~~~1/46  \\ 
 ~~~~~130 &~~~60  &~~~ .14  &~~~ .12 &~~~1/5&~~~1/46  \\ 
 ~~~~~161 &~~~50  &~~~ .43  &~~~ .57 &~~~2/13&~~~1/58  \\ 
 ~~~~~161 &~~~55  &~~~ .35  &~~~ .46 &~~~2/13&~~~1/58  \\ 
 ~~~~~161 &~~~60  &~~~ .27  &~~~ .35 &~~~2/13&~~~1/58  \\ 
 ~~~~~172 &~~~50  &~~~ .38  &~~~ .55 &~~~1/7&~~~1/61  \\ 
 ~~~~~172 &~~~55  &~~~ .32  &~~~ .46  &~~~1/7&~~~1/61 \\ 
 ~~~~~172 &~~~60  &~~~ .26  &~~~ .37 &~~~1/7&~~~1/61  \\ 
\hline
\hline\\
\end{tabular}
\end{center}
While the quoted cross-section for the initial runs is significantly
higher than what our model predicts,the cross-sections obtained from
the higher energy runs are quite consistent with our values.
Considering that detector efficiencies further reduce the expected number
of events, our $R_{2b}$ and $R_{4b}$ reproduce the observed features b) and
c) reasonably well.
As far as $\bar c s \bar s c $ is concerned ,more work is needed before 
we can confront d) with our model. 

Another concept measured for the initial runs was the rapidity 
weighted charge separation,$(\Delta Q )^y $ ; the result quoted is
0.64 $\pm$ 0.09 while standard processes give 0.38 for $u\bar uu\bar u$
and 0.38 for $q\bar q gg$.A crude estimate for this quantity can be made
for our model using Table 6 of \cite{aleph1} which gives 0.48.

The Gaussian fit to the data in \cite{rag}
gives
a width of 1.5-2.5 Gev for the dijet mass-sum distribution.At this stage it is not clear
how much of this can be attributed to the natural width of the scalars.
In our model$\Gamma^{\tilde\sigma,\tilde\pi^0}_{\bar q q} =
{g_y^2\over 4\pi}~{3M_{\tilde\sigma}\over
2}$ and
$\Gamma^{\tilde\pi^+\tilde\pi^-}_{\bar q q} =
{g_y^2\over 4\pi}~3M_{\tilde\sigma}$.
At $\rho = 1/36$ ,the former amounts to about
0.375 Gev/flavour,and with a top mass
of 170 Gev,only five flavours
contribute to the decay of
neutrals,leading to a width of
about 1.88 Gev.
Likewise, the latter works out to 0.75 Gev/generation,and 
the width of charged scalars is
about 1.5 Gev.

\twocolumn[\hsize\textwidth\columnwidth\hsize\csname
  @twocolumnfalse\endcsname

\begin{table}
\caption{COMPARSION}
\vspace{0.1in}

\begin{tabular}{lllll}
      & SUPERSYMMETRY& TWO HIGGS    & OUR MODEL & EXPERIMENT\\ 
\hline\\
Additional      &Charginos,Squarks& $ HA, H^+H^-$ & $\tilde\sigma,\vec{\tilde\pi}$   &       \\
Particles      & Neutralino...    &                &              &          \\
\hline
Additional Parameters & many & 2 Yukawa,5 Self  & 1 Yukawa,2 Self &        \\
\hline
$\sigma_{tot}$(pb)(Lower $\sqrt s$) & 1 to 7  & $\sim 1 $ & $\sim 1.17 $ &  3.5 $\pm 1.7 $   \\ 
(Higher $\sqrt s)$&similar&similar &$\sim 1 $ &  1.5 $\pm 0.8 $   \\ 
\hline
Width Of   & Large & Depends on & 1 to 2 Gev & 1.6 Gev from \\
Mass Dist  &          & Yukawa coupling &    &     curve \\
\hline
Final State& Expected & Expected& No leptons &  No leptons \\
 Leptons  &           &                 &    &   seen \\
\hline
$R_{2b}$ &  Expected &   Predominant & 1/5-1/7 with 100\% eff& 1/12 \\
\hline
$R_{4b}$ &  Expected &   Predominant & 1/45-1/63 & none seen \\
\hline
FCNC     & Fine tuning & Fine Tuning & Naturally fulfilled  &    Demanded \\
\end{tabular}
\end{table}
\vskip 2pc] 

While all these features are consistent with our model,the experiment 
seems to indicate that the invariant masses of the 
dijets to be different \cite{aleph1} while our theory predicts them 
to be degenerate. 
Also it was claimed that 
angular 
distributions do not favour
colour singlet scalars.
In our theory the dijets are colour singlet scalar or pseudoscalar.

{\bf Other explanations}:
soon after the ALEPH reports a number of papers have appeared proposing
supersymmetric candidates for the excess events \cite{susy}.
It is not possible to
give a detailed comparison between these proposals and ours but some
generic remarks can be made about these proposals. The negative outcome
of the direct searches for SUSY particles rules out many of these
explanations except the ones invoking R-parity violation. But most
such models predict more than 4 jets,leptons in final states besides
requiring fine tuning to keep the FCNC problem as well as the limits on
proton decay under control. We shall present a detailed comparison of
these proposals elsewhere but merely outline such a comparison in Table 1.

We therefore feel that the scenario presented here is quite
promising. It also has many striking predictions 
for $e^-e^+$ , $e-p$ as well as $p \bar p$
collisions  which will be
elaborated elsewhere\cite{long}.With ALEPH hoping to collect 100 $pb^{-1}$
of data at 172 GeV, we can look forward to a vindication of the model presented
here.

{\bf Acknowledgements}\\
V. Soni acknowledges the research support from the University Grants
Commission. We acknowledge many useful discussions with 
K.Abe, Tauchi, Matsui, 
P.M. Zerwas ,K. Hagiwara, G. Rajasekaran, R. Godbole,Probir Roy,D.P. Roy,Debajyoti
Choudhury,
H.S. Sharatchandra, R. Anishetty and R. Basu.

\end{document}